\documentclass[prd, aps, superscriptaddress, twocolumn, preprintnumbers, floatfix, nofootinbib]{revtex4}

\usepackage{epsf}
\usepackage{amsmath}
\usepackage{amssymb}

\usepackage{graphicx}
\usepackage{dcolumn}
\usepackage{bm}

\def\be{\begin{equation}}
\def\ee{\end{equation}}
\def\bea{\begin{eqnarray}}
\def\eea{\end{eqnarray}}
\def\cH{\cal{H}}

\usepackage{color}

\begin{document}



\title{Nonsingular Cosmology from an Unstable Higgs Field}

\author{Robert H. Brandenberger}

\email{rhb@physics.mcgill.ca}

\affiliation{Department of Physics, McGill University, Montr\'eal, QC, H3A 2T8, Canada}

\author{Yi-Fu Cai}

\email{yifucai@physics.mcgill.ca}

\affiliation{Department of Physics, McGill University, Montr\'eal, QC, H3A 2T8, Canada}

\affiliation{Department of Astronomy, Key Laboratory for Researches in Galaxies and Cosmology, \\
University of Science and Technology of China, Hefei, Anhui, 230026, China}

\author{Youping Wan}

\email{wanyp@ihep.ac.cn}

\affiliation{Theoretical Physics Division, Institute of High Energy Physics,\\
Chinese Academy of Sciences, P.O.Box 918-4, Beijing 100049, P.R.China}

\author{Xinmin Zhang}

\email{xmzhang@ihep.ac.cn}

\affiliation{Theoretical Physics Division, Institute of High Energy Physics,\\
Chinese Academy of Sciences, P.O.Box 918-4, Beijing 100049, P.R.China}


\begin{abstract}
The observed value of the Higgs mass indicates an instability of the Higgs scalar at large energy scales, and hence also at large field values. In the context of early universe cosmology, this is often considered to lead to problems. Here we point out that we can use the instability of the Higgs field to generate an Ekpyrotic phase of contraction. In the context of string theory it is possible that at very high energy densities extra states become massless, leading to an S-brane which leads to the transition between a contracting phase in the past and the current expanding phase. Thus, the Higgs field can be used to generate a non-singular bouncing cosmology in which the anisotropy problem of usual bouncing scenarios is mitigated.
\end{abstract}

\maketitle

\newcommand{\eq}[2]{\begin{equation}\label{#1}{#2}\end{equation}}

\section{Introduction}

Based on the observed value of the Higgs mass \cite{Higgs}, the Higgs potential is unstable at large energy scales where the quartic self coupling constant and hence the potential become negative \cite{instab}, with important consequences for cosmology (see e.g. \cite{Riotto1, Nima}). Since large field values lead to large energy densities, the instability will arise at large field values which are relevant to early universe cosmology. It is usually assumed that quantum gravity effects will cause the potential to turn positive again at very high values (conservatively speaking, at value where the Higgs energy density approaches the Planck density). There will be a global minimum of the potential which corresponds to Anti-de-Sitter space \footnote{An easy way to avoid this instability problem is to assume that there is new physics at scales lower than the inferred Higgs instability scale which uplifts the potential. In this paper we shall assume that this is not the case.}.

This instability is not a problem for the Standard Model of particle physics at the present time, since the usual Higgs vacua remain local vacua, and their life time is much larger than the age of the Universe. However, when applied to early universe cosmology some problems may arise. In the radiation phase of Standard Big Bang cosmology, finite temperature corrections to the potential are able to stabilize the Higgs field and localize it in the usual vacua. The Standard Big Bang Model phase of cosmology, on the other hand, must be preceded by an early phase which solves the horizon and flatness problems and allows a causal generation mechanism of fluctuations. The current paradigm for this early phase is inflation, an early phase of accelerated expansion \cite{Guth}. In simple models of inflation, the energy scale at which inflation takes place is of the order of $10^{16} {\rm GeV}$. Unless the Higgs field is directly coupled to the inflaton, the field which generates inflation,
the Higgs will perform a random walk during the period of inflation with a typical amplitude of $H$, and time scale of $H^{-1}$, where $H$ is the Hubble expansion rate during inflation. This random walk will lead to the Higgs field having a finite probability of landing in its true Anti-de-Sitter (AdS) vacuum. This is a serious problem for our universe \cite{Riotto2} (see also \cite{others} for earlier discussions of the implications of the Higgs instability for inflation, and \cite{others2} for related work).

In this paper we point out that, in the context to alternatives to inflation \footnote{See \cite{RHBrev} for a review of various alternatives to cosmological inflation.} the instability of the Higgs potential may be a virtue rather than a problem \footnote{See also \cite{Steinh} where the instability of the Higgs was used to construct the contracting phase preceding a bounce.}. Specifically, we consider a contracting universe described by General Relativity coupled to Standard Model matter. Early in the contracting phase, the equation of state will be that of cold matter, followed by a radiation-dominated phase. Eventually, the instability of the Higgs potential will set in. As we show, this generates a phase of Ekpyrotic contraction, a period which smooths out any pre-existing anisotropies. At Planck densities we postulate that a new set of fields becomes effectively massless, an effect which leads to an ``s-brane'' in the low energy effective action. This s-brane leads to a non-singular transition from contraction to expansion, as studied in detail in a different context in \cite{Costas}. The Higgs field itself bounces off a potential barrier and returns back to small field values as the universe expands.

Thus, we show that the instability of the Higgs field allows us to construct a ``natural'' non-singular cosmology which is free from the usual BKL instability \cite{BKL} which faces bouncing cosmological models \cite{Peter}. Since the universe starts in a matter phase of contraction, our model yields a simple realization of the ``matter bounce'' alternative to inflation \cite{FB, Wands}.

\section{Model}

We will be considering the Higgs sector of the Standard Model of
particle physics. The Lagrangian is
\be
{\cal L}_0 \, = \, X - V(h) \, ,
\ee
where $h$ denotes the Higgs field,
$X$ is the standard kinetic term
($X = \frac{1}{2} (\partial_\mu h)^2$) and $V(h)$ is the
Higgs potential energy. The bare potential energy is given by
\be
V(h) \, = \, \frac{1}{4} \lambda_0 (h^2 - v^2)^2 \, ,
\ee
where $\lambda_0$ is the tree level renormalized coupling constant
and $v$ is the vacuum expectation value of the field.

The bare potential is subject to quantum corrections. At one
loop level, the corrections have been studied in detail in
\cite{Sher} (see also \cite{RHBRMP}). There are logarithmic
corrections to the
potentail which enter with a positive sign for bosons and
with a negative sign for fermions. For the observed value
of the top quark the fermion contribution dominates
and the one loop effective potentail can be written as
\be
V^{(1)}(h) \, = \, \frac{1}{4} \lambda(h) (h^2 - v^2)^2 \, ,
\ee
with
\be \label{effLambda}
\lambda(h) \, = \, \lambda_0 - b {\rm ln}\bigl( \frac{h^2}{\Lambda^2} \bigr) \, ,
\ee
where $b$ is a positive number set by the top quark Yukawa
coupling constant, and $\Lambda$ gives the scale at which the
instability sets in, which according to the current best Higgs and top
quark mass measurements is about $10^{10} {\rm GeV}$ \cite{instab}.
In the following we shall work with the
one-loop effective potential but we will omit the superscript.
In our numerical study we will work with a larger value of $\Lambda$
in order to reduce the magnitude of the hierarchy in energy and
time scales.

At field values which correspond to Planck-scale or string scale
energy densities we assume that there will be corrections which
uplift the potential. We consider (following \cite{Nima})
the leading order non-renormalizable term
\be \label{extra}
\delta V \, = \, g  \frac{h^6}{M^2}  \, ,
\ee
where the coupling constant $g$ can be absorbed into the
mass scale $M$ and we can set $g = 1$. In Figure 1 we
present a sketch of the one loop effective potential
including the above term.

\begin{widetext}

\begin{figure}[t]

\includegraphics[height=8cm]{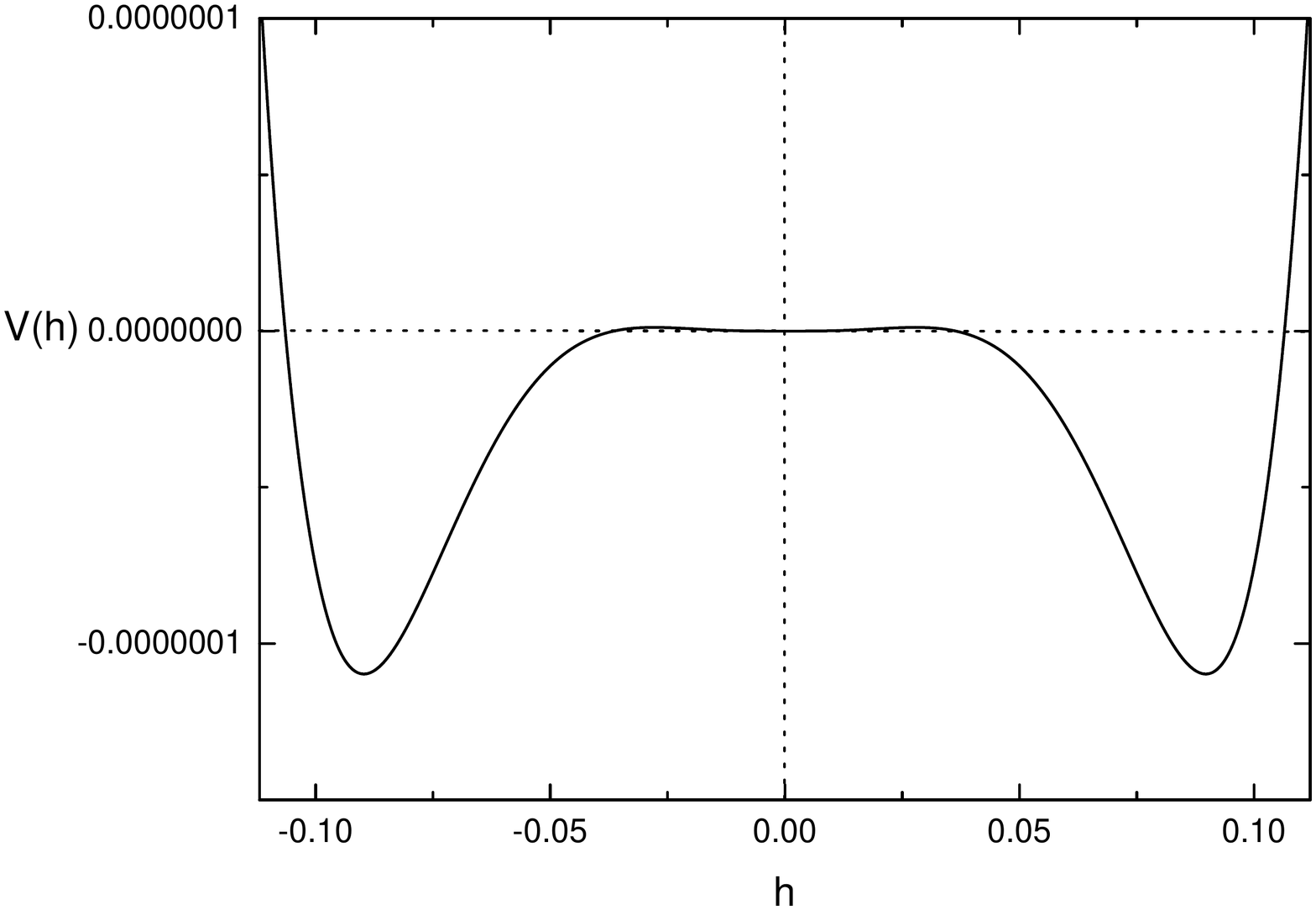}

\caption{Sketch of the one loop effective potential of the Higgs field
including the extra term (\ref{extra}) coming from postulated
quantum gravity effects. The instability of the Higgs sets in
at field values above $\Lambda$. Note that the local minimum
of $V(h)$ at $h = 0$ is not visible on this plot due to the
large hierarchy of scales between the electroweak symmetry
breaking scale and the Planck scale. We have used the
following values of the constants appearing in the potential:
$v = 246 {\rm GeV}$, $\lambda_0 = 0.129$, $b = 0.0187$,
$\Lambda = 2.4 \times 10^{15} {\rm GeV}$,
$g = 1$ and $M = 2.4 \times 10^{18} {\rm GeV}$.
The first three of these parameters reflect the measured masses of
the Higgs and the top quark, the final term reflects the
assumption of a quantum gravity-induced wall in the potential
at values of $h$ corresponding to the reduced Planck mass.} \label{fig:1}

\end{figure}

\end{widetext}

Following \cite{Costas}, we assume that at high energy
densities a new sector of states becomes effectively
massless. This follows if we consider a superstring
model which reaches an enhanced symmetry point at
some critical density \cite{beauty, Watson}. In this case
these states must be included in the low energy effective
action as a term arising only at a particular density, or
equivalently at a particular time $t_c$ when this critical
density is achieved. The action $S$ hence contains a term of
the form
\be \label{Sbrane}
\delta S \, = \, \mu \delta(t - t_c) \, .
\ee
Such a term is called an {\it S-brane}, and $\mu$ is its
tension which is set by the mass scale of the new physics.

An S-brane can be viewed as a relativistic topological defect
which is space-like. It is well known that for such defects
the pressure in direction of the defect is negative (i.e. positive
tension), and the pressure perpendicular to the defect vanishes.
For an S-brane, this implies that the induced energy density
vanished and the pressure is negative. Hence, for the model
we are considering the energy density $\rho$ and pressure $p$ are given
by
\bea
\rho \, &=& \, X + V \\
p \, &=& \, X - V - \mu \delta(t - t_c) \, .
\eea
From the above it follows that an S-brane leads to the violation
of the weak energy condition which allows for a transition
between a contracting phase and an expanding phase.

The cosmological scenario which we have in mind is now the
following. We begin in a contracting universe with the Higgs
field close to today's minimum $h = v$. Initially, $h$ is oscillating
about $v$ with an amplitude which is small compared to $v$.
The corresponding equation of state is that of (when averaged
over time) cold matter, i.e. $p = 0$. The amplitude of field
oscillations will grow as the universe contracts, and eventually
it reaches of order $v$. After that point, $h$ is free also to
explore negative field values. In fact, after some time of
contraction the local Higgs potential barrier at $h = 0$
becomes negligible and $h$ will oscillate in a potential
which looks quartic. At this point the equation of state will
change to that of (again averaged over time) radiation, i.e.
$p = \frac{1}{3} \rho$. The early dynamics of $h$ in our
model is shown in Fig. 2, as follows from numerically solving the
equations of motion discussed in the following section.

\begin{widetext}

\begin{figure}[t]

\includegraphics[height=8cm]{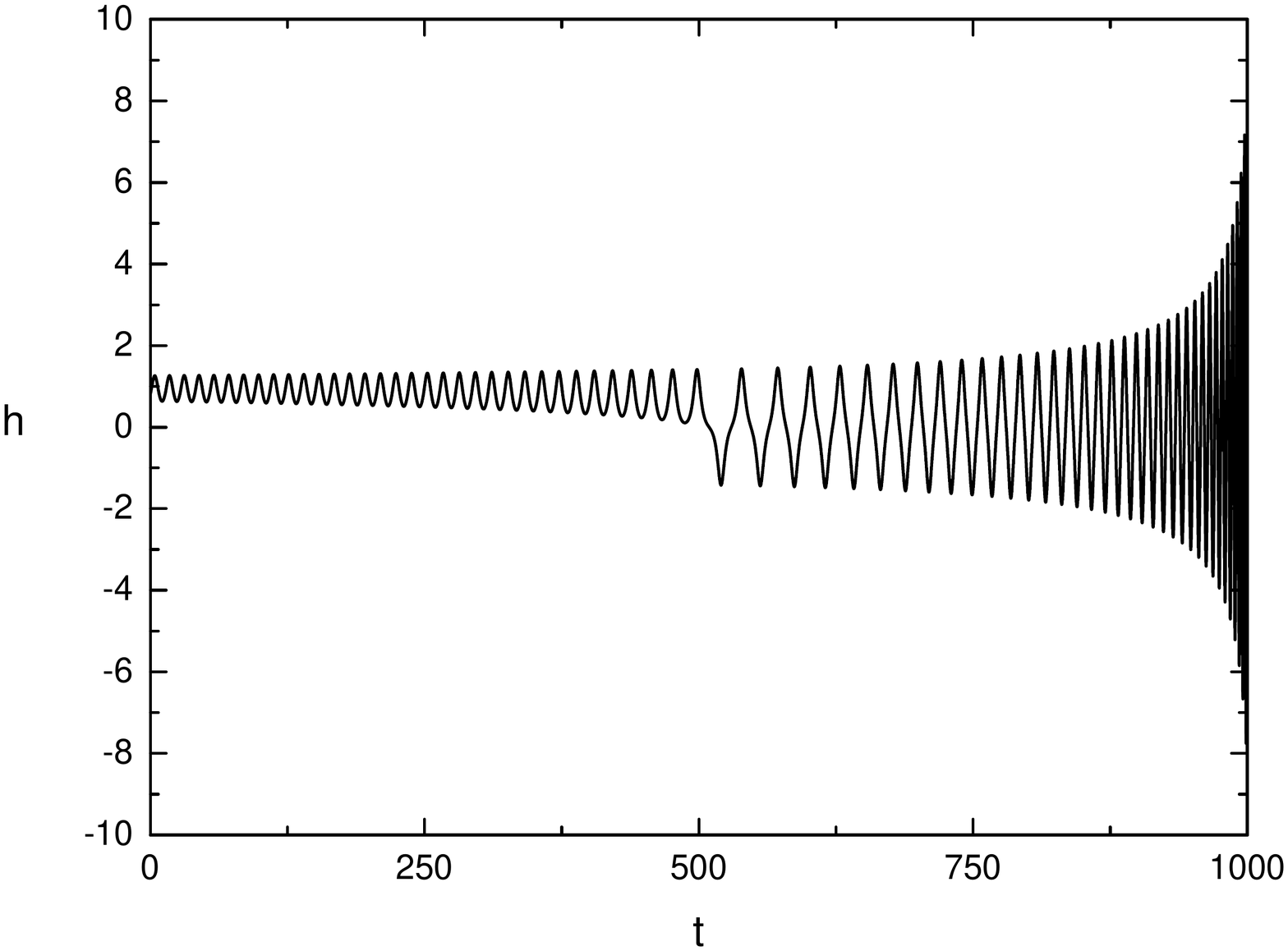}

\caption{The evolution of the Higgs field early in the contracting
phase. The initial conditions were taken to be $h = v + \delta h$
with a small offset $\delta h = 0.1 v$ and ${\dot{h}} = 0$.
Initially, $h$ oscillates about
$h = v$, but because of the Hubble antidamping the amplitude of
oscillations grows and eventually $h$ will be able to cross over
the local potential maximum at $h = 0$. Subsequently, $h$
executes oscillations about $h = 0$. Before the transition the
time-averaged equation of state is $w = 0$, afterwards
$w = 1/3$. The figure shows the evolution of $h$ (vertical
axis) as a function of time (horizontal axis). The field
values are in units of $v$, the time values in units of $v^{-1}$.
In this numerical
simulation, the $h$ dependence of the quartic coupling
constant and the extra $h^6$ terms are neglected since
they are not important during the initial dynamics. Also,
in order to see both the increase in the amplitude and the
oscillation period easily, we have set $v = 10^{-3}$ in
Planck units. } \label{fig:2}

\end{figure}

\end{widetext}

The amplitude of the (anharmonic) oscillations of $h$ will
continue to grow until it reaches the local maximum
of the potential at $h \sim \Lambda$. At that point
the instability of $h$ sets in and $h$ will start rolling
down the potential towards negative values of $V$.
Once the potential becomes negative, the equation of
state of the Higgs field becomes of {\it Ekpyrotic} type
\cite{Ekp}, i.e.
\be
w \, \equiv \, \frac{p}{\rho} \, > \, 1 \, .
\ee
As was realized in \cite{EMB} in the context of the
{\it matter bounce} scenario, this phase of Ekpyrotic
contraction has the virtue of diluting anisotropies \cite{noBKL}.
Thus, our scenario in a completely natural way solves the
main problem of a bouncing scenario, namely the
anisotropy problem.

Once the energy density in the contracting phase approaches
the string scale (or Planck scale) density, two effects take
place. Firstly, the Higgs field hits the ``potential wall'' where
$V(h)$ sharply rises due to the extra contribution $\delta V$
to the potential from (\ref{extra}). This causes $h$ to slow down
and $V(h)$ to turn positive. Secondly, at the string density the
S-brane (\ref{Sbrane}) is encountered. We will take this to
happen at the time when $h$ comes to rest, i.e. $X = 0$ (we
comment later on this assumption). At this point, a transition
between contraction and expansion takes place. Because
for a symmetric bounce the kinetic energy is negligibleimmediately
before and after the bounce the equation of state parameter
$w$ will approach $w = -1$ from both sides. At the
bounce point itself, the S-brane leads to a value of $w$ which
is formally $w = - \infty$ (this shows the consistency
of our analysis with the general theorems of \cite{Zhang} that
the single fluid matter with a crossing of $w = -1$ is required to
obtain a bounce \cite{quintom}).

After the bounce, the Higgs
field $h$ will roll back down the potential (towards the origin)
picking up kinetic energy, and then using this kinetic energy
to roll back to $h \sim \Lambda$ and back into the local
minimum at $h = v$. The dynamics of $h$ in the phase
of Higgs instability and around the S-brane bounce is shown
in Figs. 3 and 4. Once again, these plots are obtained by
numerically solving the equations of motion discussed in
the following section.

\begin{widetext}

\begin{figure}
\centering
\begin{minipage}{.5\textwidth}
  \centering
  \includegraphics[width=1.1\linewidth]{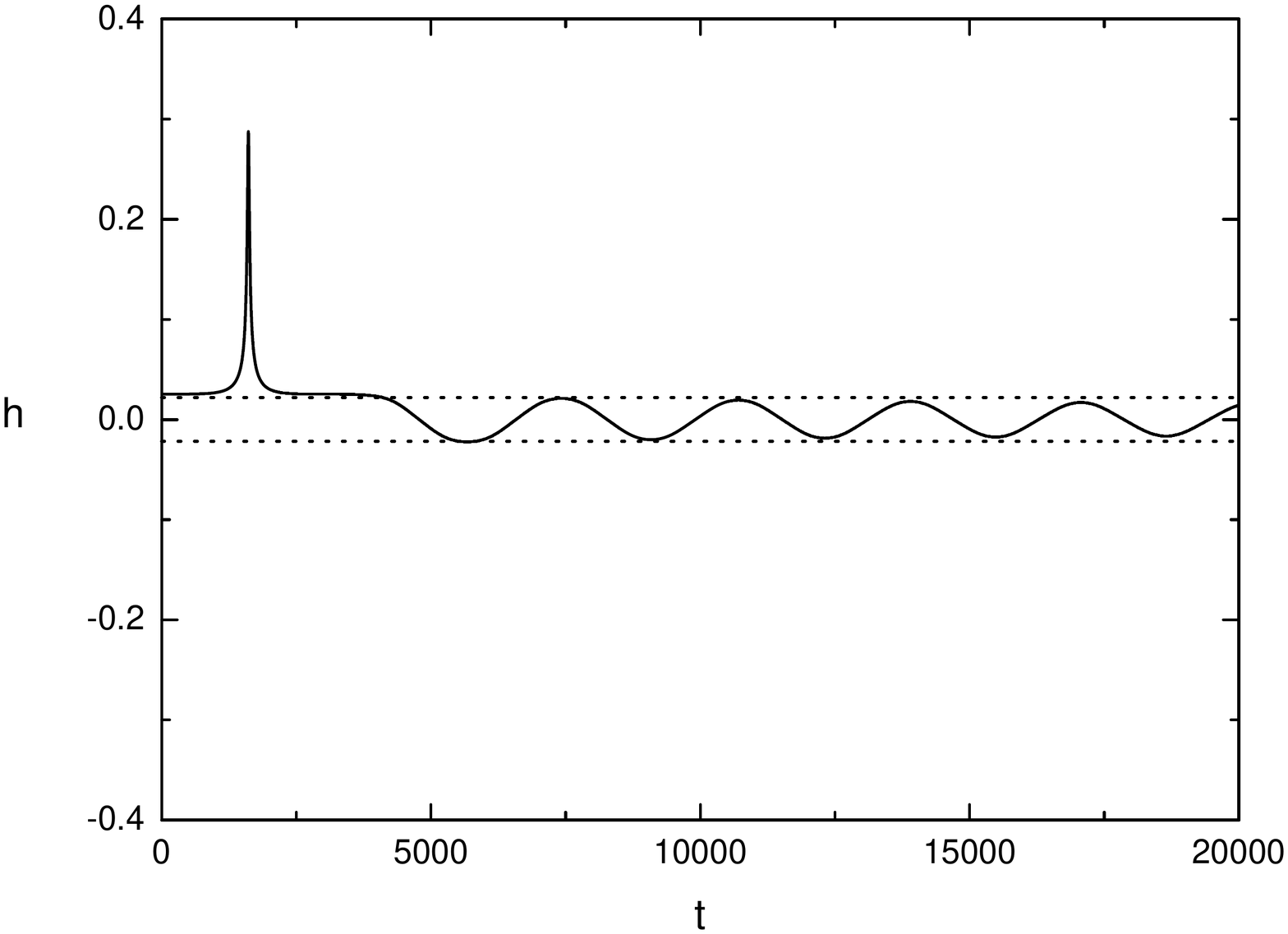}
  \label{fig:test1}
\end{minipage}%
\begin{minipage}{.5\textwidth}
  \centering
  \includegraphics[width=1.1\linewidth]{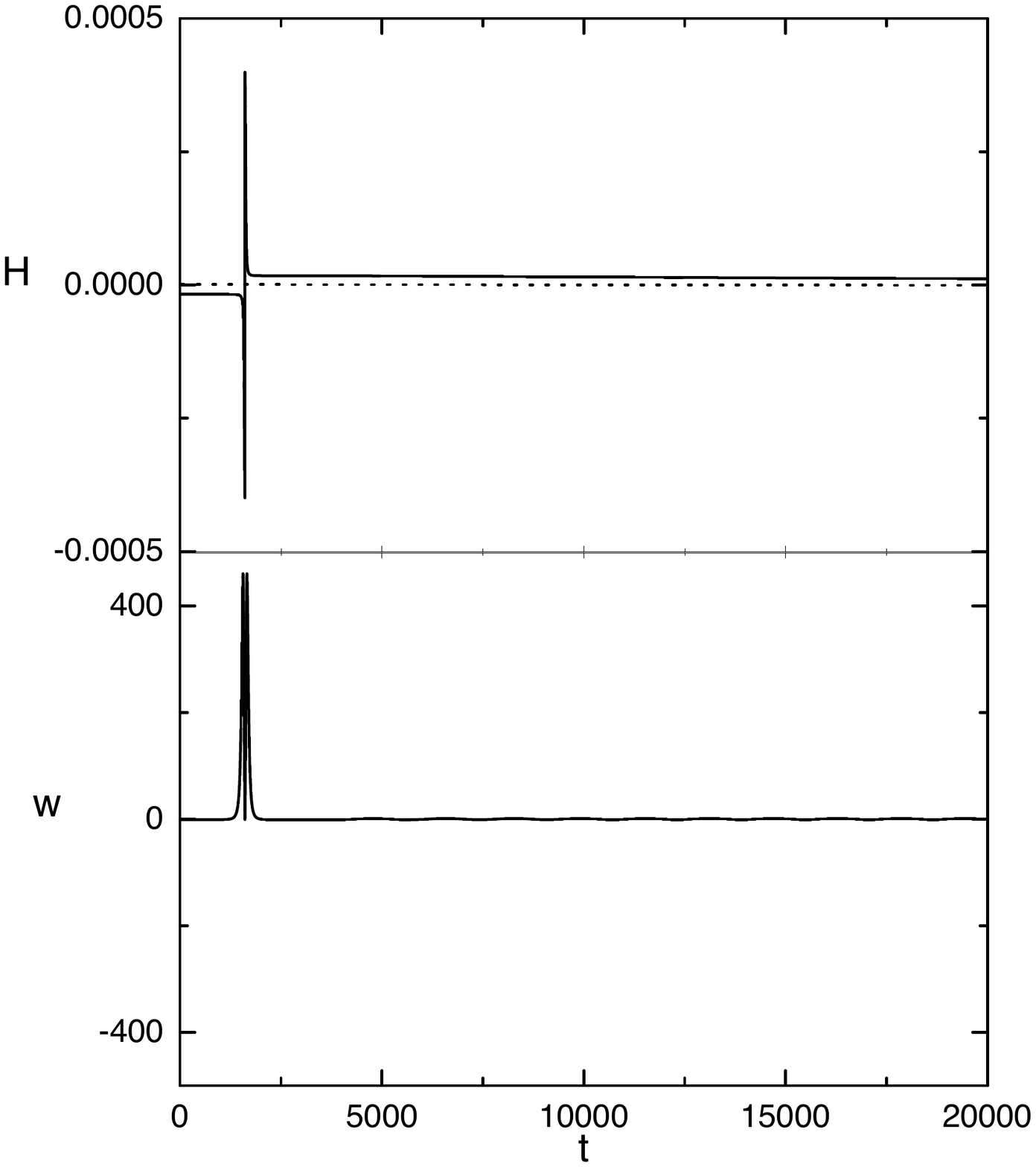}
  \label{fig:test2}
\end{minipage}

\caption{The evolution of the Higgs field in the contracting
phase after the onset of the instability. The initial conditions
were taken to be $h$ just to the left of the local maximum
of the potential with a small initial velocity sufficient to push
$h$ over the barrier (specifically $h = 2 \times 10^{-2}$ and
${\dot{h}} = 2 \times 10^{-5}$ in reduced Planck units).
The left panel shows the evolution of $h$ (vertical
axis) as a function of time (horizontal axis). The field
and time values in Planck units. To guide the eye we
have drawn horizontal dashed lines at the value of $h$
corresponding to the onset of the Higgs instability. The
numerical result shows that at late times the Higgs field
becomes localized in the metastable Higgs region
near $h = 0$.
The  right panel shows the equation of state parameter $w$
(solid curve) and the Hubble parameter (dashed curve)
(each on the vertical axis) as a function of time.
The field is seen to rapidly accelerate down the potential,
reach negative values of the potential with a resulting
Ekpyrotic equation of state $w \gg 1$. After crossing
the minimum of the potental, $h$ climbs up the steep potential
and comes to rest. At the point when $h$ comes to rest
we have inserted an S-brane chosen to yield a symmetric
bounce, i.e. $H$ simply changes sign, and the values
of $h$ and ${\dot{h}}$ do not change. As can be
seen, after the bounce $h$ falls back down to the
minimum value of its potential, climbs back up
towards $h = 0$, and ends up oscillating about
$h =0$.  At the large
field values considered here, the double well structure
of the potential near $h = 0$ has a neglible effect. Hence,
we have here taken $v = 0$.}  \label{fig:3}

\end{figure}

\begin{figure}
\centering
\begin{minipage}{.5\textwidth}
  \centering
  \includegraphics[width=1.1\linewidth]{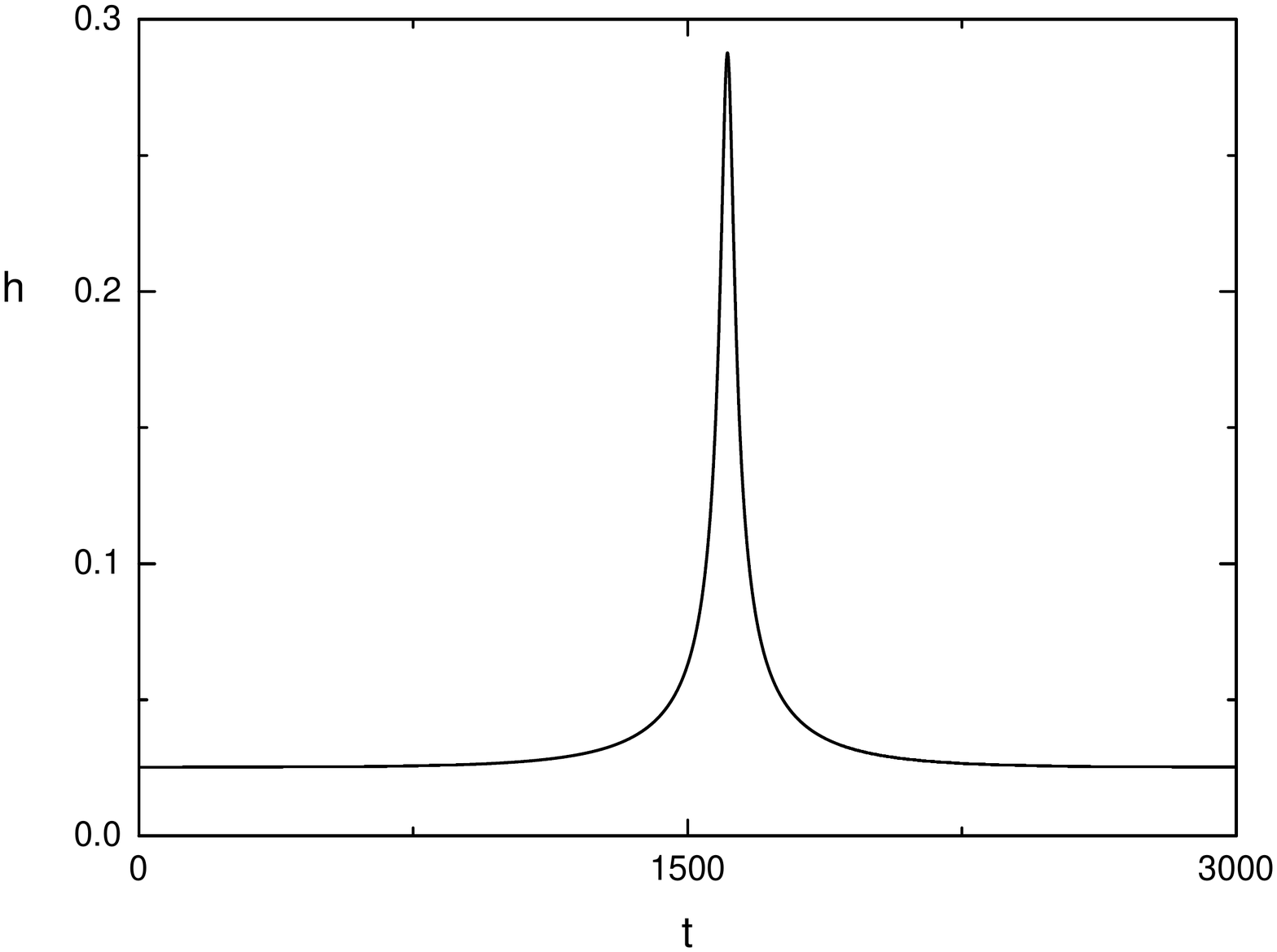}
  \label{fig:test3}
\end{minipage}%
\begin{minipage}{.5\textwidth}
  \centering
  \includegraphics[width=1.1\linewidth]{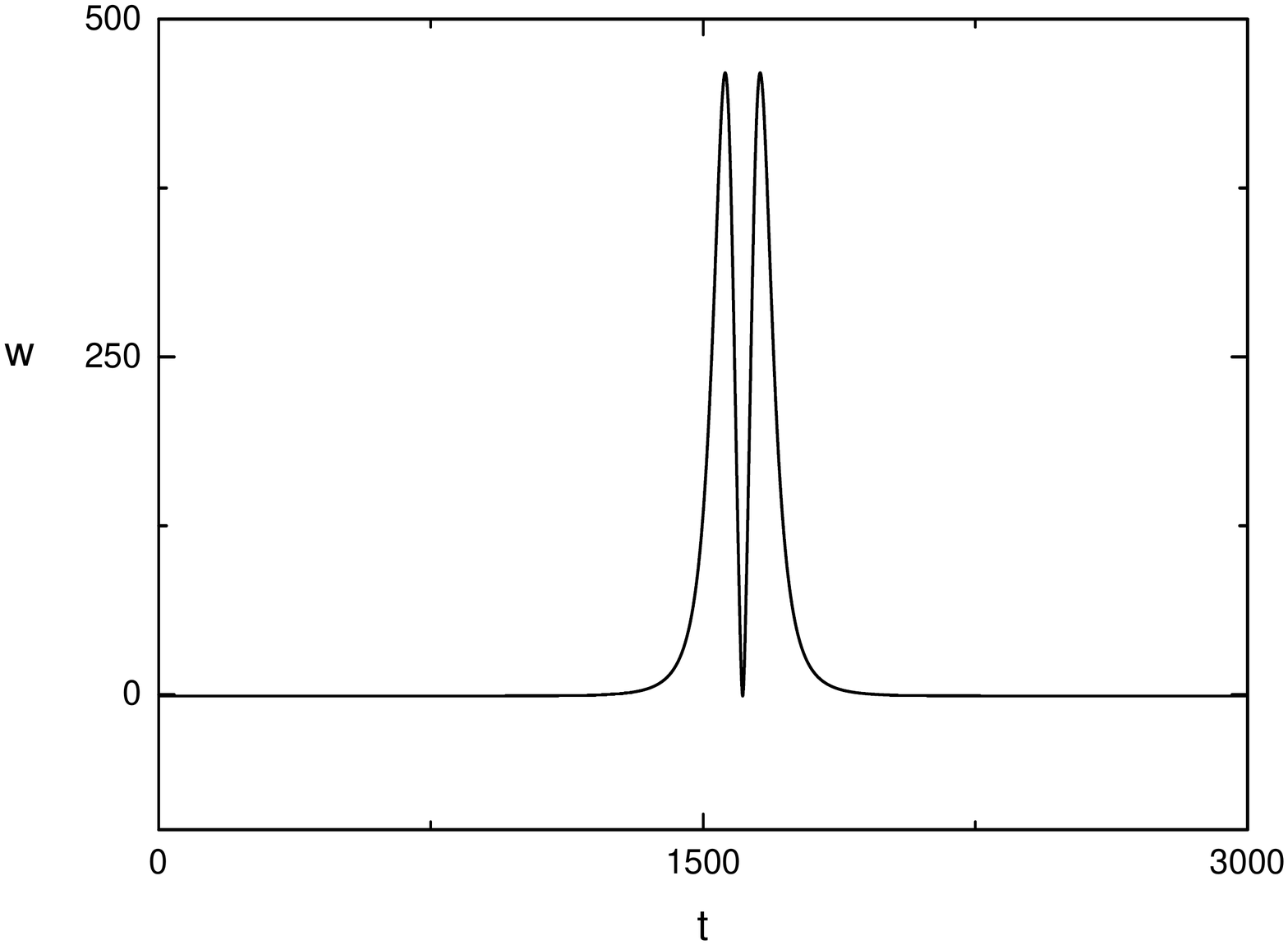}
  \label{fig:test4}
\end{minipage}

\caption{This figure is a blowup of the previous one and shows
the dynamics of $h$ and $w$ near the bounce point.
In particular, the figure shows that immediately before and
after the actual bounce point $w$ turns negative. The actual
numbers show that $w = -1$. The axes and units are as in the previous
figure. In this case we do not show the evolution of the
Hubble parameter.}  \label{fig:4}

\end{figure}

\end{widetext}

In the above we have focused on the Higgs sector of the
Standard Model matter. We should keep in mind that
the other particles of the Standard Model also contribute
to the dynamics. If we have in mind starting the contracting
phase in a state which looks like the time reverse of our
present universe, the rest of the Standard Model matter
will initially be in a cold matter-dominated state as well. The
radiative degrees of freedom of the Standard Model matter
will lead to a transition between the matter-dominated phase
and the radiation phase earlier than if we only consider
the Higgs field dynamics. This does not change our
scenario. It simply causes the Higgs field to climb its
potential towards the instability point $h \sim \Lambda$
at a different rate. Once the Higgs field starts its descent
towards negative values, the Ekpyrotic equation of state
of the Higgs field will cause the Higgs to rapdily come
to dominate matter, and from then on the analysis is
exactly how it was described above.

In the following we will take a closer look at the dynamics of
the Higgs system. Since in the Ekpyrotic phase spatial
gradient terms also get washed out we will focus on the
homogeneous dynamics.

\section{Equations of Motion}

As usual, it is convenient to write the dynamics in terms of
a rescaled field
\be
u \, \equiv \, a h \, ,
\ee
and in terms of conformal time $\eta$ which is related to
the physical time $t$ via \footnote{Note that in the contracting
phase $\eta$ is negative and approaches $\eta = 0$.}
\be
dt \, \equiv \, a(t) d\eta \, .
\ee
The Higgs field equation of motion then becomes
\be \label{hEoM}
u^{\prime \prime} - \bigl( {\cH}^2 + {\cH}^{\prime} \bigr) u \, = \,
- a^3 V_{, h} \, ,
\ee
where $\cH$ is the Hubble expansion rate in conformal time,
namely ${\cH} = a^{\prime}/a$, and a prime denotes the derivative
with respect to $\eta$.

During the radiation phase of contraction the second term on the
left hand side of (\ref{hEoM}) vanishes and, in the limit
$|h| \gg v$ the equation reduces to
\be
u^{\prime \prime} \, = \, - \lambda u^3 \, .
\ee
For field values $|h| \ll \Lambda$, i.e. far from
the Higgs instability point, we can take $\lambda$ to
be a positive constant, and hence the solutions yield anharmonic
oscillations.

Beyond the instability point the sign of $\lambda$ changes.
As mentioned before, the runaway of $h$ to values with
negative potential leads to an Ekpyrotic equation of state
$w \gg 1$. This leads to slow contraction
\be
a(t) \, \sim \, t^p \, ,
\ee
with $p \ll 1$. In this case, the terms of (\ref{hEoM}) involving
${\cH}$ and its derivatives are negligible. During a short time
interval during which the change in the logarithm in (\ref{effLambda})
can be neglected the equation of motion takes the form
\be
u^{\prime \prime} \, = {\tilde{\lambda}} u^3 \, ,
\ee
where ${\tilde{\lambda}} > 0$, where has rapidly growing solutions of
the form
\be
u(\eta) \, = \, \frac{f}{\eta} \, ,
\ee
with
\be
f^2 \, = \frac{2}{{\tilde{\lambda}}} \, .
\ee

The above rapid growth of $h$ leads to the potential energy which scales
as
\be
V \, \sim \, - \frac{2 f^2}{\eta^4} a^{-4}
\ee
which show that the energy density in the Higgs field rapidly comes
to dominate over all other forms of energy density (in particular that
of regular radiation which scales as $a^{-4}$).

The system of equations is completed with the Friedmann-Robertson-Walker
equations which take on a simpler form in terms of physical time.
During the stages when the Higgs field dominates the energy-momentum
tensor of matter these equations are
\be
H^2 \, = \, \frac{8 \pi G}{3} \bigl( X + V \bigr)
\ee
where $H \equiv {\dot{a}}/a$ is the Hubble expansion rate in
terms of physical time, and $G$ is Newton's gravitational
constant, and
\be
\frac{\ddot{a}}{a} \, = \, - \frac{4 \pi G}{3} \bigl( 4X - 2V - 3 \mu \delta(t - t_c) \bigr) \, .
\ee
More specifically, inserting the expressions for the equation for the change in $H$ becomes
\be \label{HdotEq}
\dot{H} \, = \, - 8 \pi G X + 4 \pi G \mu \delta(t - t_c) \, ,
\ee
where $t_c$ is the time when the energy density has increased to the point
that the S-brane appears.

From (\ref{HdotEq}) it follows that once the S-brane is hit, the value of the Hubble
constant jumps by an amount
\be \label{Hchange}
\Delta H \, = \, 4 \pi G \mu \, .
\ee
In order to obtain a cosmological bounce this jump in $H$ has to be large
enough to change the sign of $H$. Let us estimate the numbers. For concreteness
let us assume that the S-brane is due to string states becoming massless, as in
the scenario of \cite{Costas}. In this case the value of $\mu$ is given by
dimensional analysis by
\be
\mu t_s^{-1} \, = m_s^4 \, ,
\ee
where $m_s$ is the string scale and $t_s$ is the associated time, namely
\be
t_s \, = \, G^{-1/2} m_s^{-2} \, .
\ee
Hence
\be
\mu \, = \, G^{-1/2} m_s^2 \, .
\ee
This value has to be compared with the value of $H$ at the time when
the Higgs energy density reaches the string scale density $m_s^4$.
By the first Friedmann equation this is given by
\be
H \, \sim G^{1/2} m_s^2 \, .
\ee
Thus, we see that the expected jump in $H$ due to the S-brane is
large enough to change contraction into expansion.

Whether the cosmological bounce is symmetric or not appears
to depend on details of the construction. If the time $t_c$ is the
time when the Higgs field comes to rest when running up the
potential wall as stringy values, then the bounce will be
symmetric. If the time $t_c$ arises earlier, then after the
bounce the Higgs field will continue to roll up the hill for a while
before turning around. Since the universe keeps contracting
until the S-brane is hit, even if $h$ has already turned around,
it is also possible for the bounce to occur after $h$ has already
reached its maximal value.

Let us take a closer look at the matching conditions across the
S-brane. Given the change of $H$ across the brane by the
amount (\ref{Hchange}), the values of $h$ and ${\dot{h}}$
across the brane are also determined. We will assume that
$h$ does not jump. In this case, the equations
\bea
H_{-}^2 \, &=& \, \frac{8 \pi G}{3} \bigl[ V_{-} + X_{-} \bigr] \, , \\
H_{+}^2 \, &=& \, \frac{8 \pi G}{3} \bigl[ V_{+} + X_{+} \bigr] \,
= \, \bigl( H_{-} + \Delta H \bigr)^2 \, ,
\nonumber
\eea
where the subscripts $-$ and $+$ indicate the quantities
right before and right after the bounce. These equations
determine the value of ${\dot{h}}$ after the bounce. For
a symmetric bounce with $X_{-} = 0$ and $H_{+} = - H_{-}$
we obtain $X_{+} = 0$, i.e. $h$ starts at rest after the
bounce.

In the numerical simulation whose results are presented
in Figs. 3 and 4 we have solved the equations of motion
without the brane for times before the bounce is reached.
We assume a symmetric bounce, i.e. that the bounce occurs
when $h$ has climbed up the potential at large field values
to positive values and comes to rest. At this point, we
reverse the sign of the value of the Hubble parameter, as
discussed above. After the bounce we again solve the
equations of motion without the brane. Note that at the
bounce point $w = -1$ since $X = 0$.

The figures show that the scenario argued for in the
above approximate analytical considerations is indeed
obtained. In particular, there is an Ekpyrotic phase of
contraction followed by a non-singular bounce, and at
late times in the expanding phase the Higgs field ends
up in its ``regular'' vacuum state with $|h| = v$. A
blowup of Figure 4 shows that for the value of $\Lambda$
chosen, $\Lambda = 10^{-3}$ in Planck units, the Ekpyrotic
phase of contraction lasts for about $300$ Planck times.
We have verified numerically that the length of the
period of Ekpyrotic contraction scales roughly as
$\Lambda^{-1}$. Thus, for the value $\Lambda \sim 10^{-7}$
in Planck units indicated by the current measurements of
the Higgs and top quark mass, we obtain about
$3 \times 10^{6}$ Planck times of Ekpyrotic contraction.
Whether this is sufficient to completely solve the
anisotropy problem depends on the initial anisotropy
at the beginning of the evolution.

 On the other hand, the tuning on the parameter $\mu$
 required to obtain a sufficiently symmetric bounce
 to allow the Higgs field to relax to $|h| = v$ at late
 times becomes more severe the smaller $\Lambda$
 is.

\section{Cosmological Scenario}

The cosmological scenario we have developed is the following.
We start in a matter-dominated phase of contraction with the
Higgs field close to its current vacuum value $v$. After some
amount of contraction there will be a smooth transition to
a radiation phase of contaction. During both periods $h$
will be oscillating (initially about $v$ and later about $0$).
Once the amplitude of oscillation reached a value of order
$\Lambda$, the Higgs instability will set in. The Higgs field
rolls off to negative values of the potential, leading to an
Ekpyrotic phase of contraction during which the energy
the Higgs energy density comes to dwarf the energy
density in all other forms of matter, and also smooths
out anisotropies. Eventually the energy density increases
to the point when the dynamics hits an S-brane, at which
point a non-singular transition from contraction to expansion
sets in. In the expanding phase the Higgs field re-traces its
evolution in the contracting phase, eventually landing it
back in its vacuum state $h = v$.

This scenario provides a realization of the {\it matter bounce}
alternative to cosmological inflation as a theory for the
origin of structure in the universe, and as a solution of the
horizon problem. According to this scenario, we start early
in the contracting phase with fluctuations in their quantum
vacuum state. The growth of the curvature fluctuations
on super-Hubble scales in the contracting phase then
transforms the spectrum from a vacuum spectrum to
a scale-invariant one on length scales which exit the
Hubble radius during the matter phase of contraction
\cite{FB, Wands}.

The key question for late time cosmology is whether the
scale-invariance of the spectrum of curvature fluctuations
survives the bounce. This question has been studied
in many toy models on non-singular bouncing cosmology,
e.g. in the quintom bounce \cite{quintom, quintomflucts}, in the
Horava-Lifshitz gravity bonce \cite{Elisa}, the bounce \cite{HLbounce}
obtained in Horava-Lifshitz gravity \cite{HL} and in the
ghost condensate and Galileon bounces \cite{Francis} (see
\cite{Chunshan, Damien} for the background models),
and it was found that on length scales larger than the time
scale of the bounce phase the spectral shape is unchanged.
The issue is more subtle, however, in the case of an
S-brane bounce, as discussed in detail in \cite{Durrer}.
In this case, the final spectrum depends sensitively on
the coordinate system in which the brane is defined.
In the case of purely adiabatic fluctuations the scale-invariance
of the spectrum is preserved.

\section{Discussion}

We have shown that the instability of the Higgs potential can have
positive consequences for early universe cosmology when considered
not in the context of the inflationary scenario, but in the context of
a bouncing scenario. More specifically, we have shown that the
instability of the Higgs field generates an Ekpyrotic phase of
contraction which smooths out anisotropies and hence solves the
key problem facing bouncing cosmologies.

We have argued that an S-brane arising from stringy effects can
lead to a non-singular bounce. When the S-brane is considered
to have infinitesimal thickness in temporal direction there will be
a jump in the value of the Hubble constant, but if the S-brane
is smeared out in the same way that topological defects are (they
have finite thickness) then the cosmological evolution is
completely non-singular.

Our scenario hence gives a realization of the matter bounce
alternative to the inflationary scenario of structure formation.

Note that to obtain a symmetric bounce a certain amount of tuning
of $\mu$ is required. For asymmetric bounces the danger is that
$h$ will not be able to relax to $h = \pm v$ at late times in the
expanding phase, but comes to rest in a AdS minimum.

\acknowledgements{Two of us (RB and YC) wish to thank X. Zhang and the Theory Division of the Institute of High Energy Physics in Beijing for hospitality during the time which this work was started. RB is supported by an NSERC Discovery Grant, and by funds from the Canada Research Chair program. YFC is supported in part by the National Youth Thousand Talents Program and by the USTC start-up funding under Grant No. KY2030000049. Y. W. and X. Z. are supported in part by the National Science Foundation of China under Grants No. 11121092, No.11375202 and No. 11033005. The research is also supported by the Strategic Priority Research Program, the Emergence of Cosmological Structures of Chinese Academy of Sciences, Grant No. XDB09000000. We also wish to thank Fa Peng Huang for comments on our manuscript.}

\end{document}